# The Novel ABALONE Photosensor Technology – 4-Year Long Tests of Vacuum Integrity, Internal Pumping and Afterpulsing

Daniel Ferenc[a], Andrew Chang[a], and Marija Šegedin Ferenc[b]

(a) Physics Department, University of California, Davis
One Shields Ave., Davis, CA-95616, USA
(b) PhotonLab, Inc., 3315 Oyster Bay Ave., Davis, CA-95616, USA

ABSTRACT

The ABALONE Photosensor Technology (U.S. Patent 9,064,678(2015)) has the capability of supplanting the expensive 80-year-old Photomultiplier Tube (PMT) manufacture by providing a modern and cost-effective alternative product. An ABALONE Photosensor comprises only three monolithic glass components, sealed together by our new thin-film 'adhesive.' In 2013, we left one of the early ABALONE Photosensor prototypes intact for continuous stress testing, and here we report its long-term vacuum integrity. The exceptionally low ion afterpulsing rate (approximately two orders of magnitude lower than in PMTs) has been constantly improving. We explain the physical and technological reasons for this achievement. Due to the cost-effectiveness and the specific combination of features, including low level of radioactivity, integration into large-area panels, and robustness, this technology can open new horizons in the fields of fundamental physics, functional medical imaging, and nuclear security.

1. INTRODUCTION

The ABALONE Photosensor Technology [1] has the capability of supplanting the expensive 80-year-old Photomultiplier Tube (PMT) manufacture by providing a modern and cost-effective alternative product, and thus opening new horizons in all application areas, including large experiments in fundamental physics as well as new types of scanners for both functional medical imaging and nuclear security [2].

To this day, the broad field of large-area photon detection has relied on the expensive (>$100,000/m$^2$) vacuum PMT manufacture (Fig.1, Fig.2(a,b)). Furthermore, the millimeter-sized all-silicon devices, particularly the Geiger-Mode Avalanche Photo Diodes (G-APDs) (most often

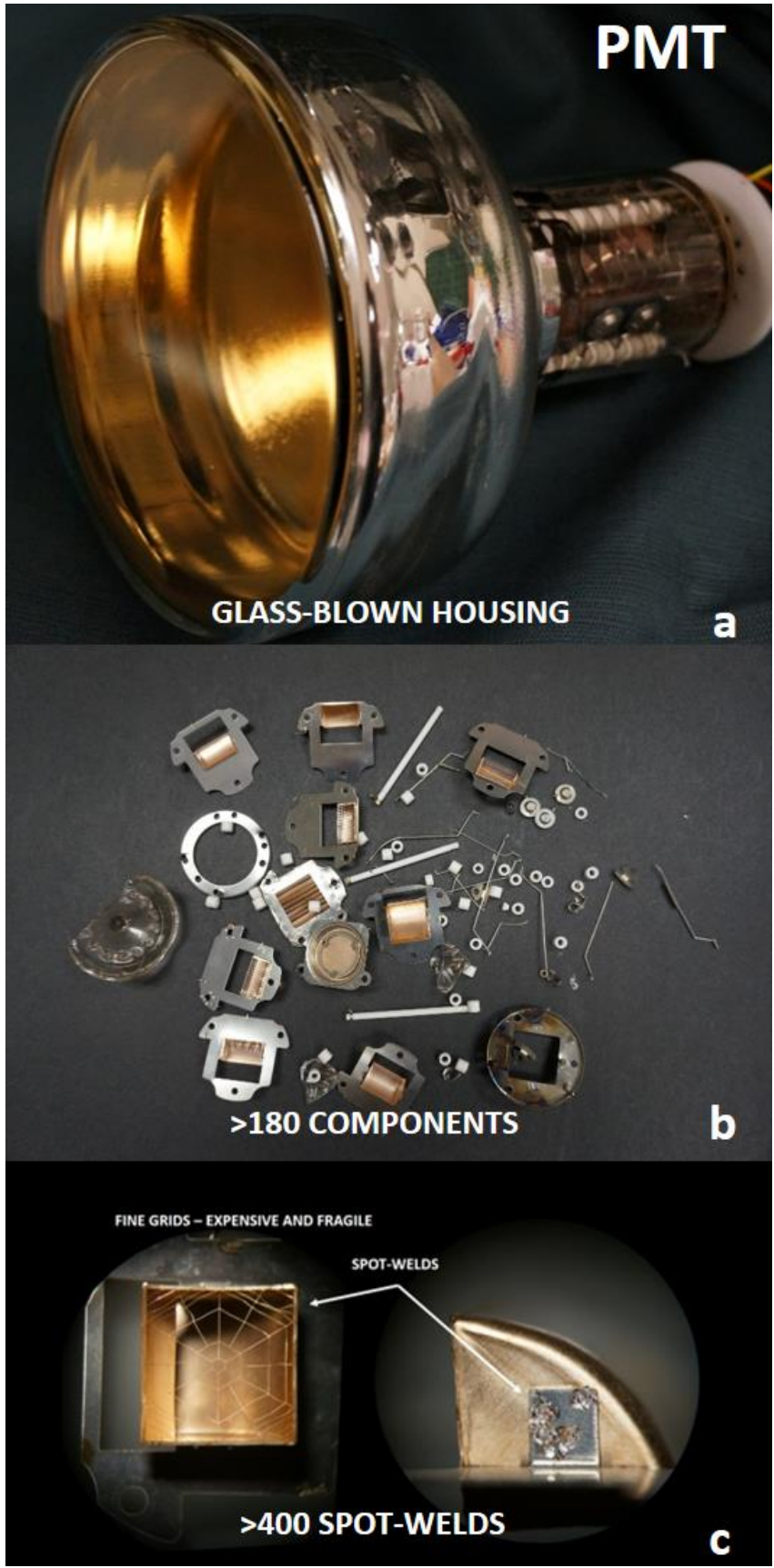

inaccurately called Silicon Photo Multipliers[1] (SiPMs)) are and will remain far too expensive (currently ~$10 million/m$^2$) for truly large-area applications.

Fig.1 The anatomy of a Photomultiplier Tube (PMT). The reasons why the 80-year old PMT manufacture could not have evolved into a modern cost-effective technology include the large number of hand-made components (a, b) and the labor-intensive assembly methods (c). However, the real, insurmountable technological limitation lies in the intrinsic combination of materials that precludes modern continuous production-line assembly.

In contrast to the manufacturers of PMTs, many modern industries that are also based on vacuum processing technologies have been thriving with the production of numerous items, ranging in complexity from blank compact discs (CDs) (<$10/m$^2$) to plasma TV screens (~$500/m$^2$). Our studies led to the conclusion that the reason why the PMT manufacture could never have evolved into such a modern, cost-effective vacuum technology is simply due to the combination of materials that the PMTs are made of (Fig.1, Fig.2(a,b)), which is intrinsic to its very concept. For the reasons discussed below, this presents an insurmountable technological limitation.

For the PMTs, the combination of materials necessitates a daylong static vacuum bakeout for the cleaning of the components, which precludes any form of continuous-line production. Any bakeout within a pumped vacuum environment is a statistical process, characterized by a non-linear trade-off between the temperature and the bakeout time. The presence of glass among the inseparable PMT components limits the maximum bakeout temperature to far below the optimal outgassing temperatures for metals. Therefore, the trade-off must occur at the expense of bakeout time.

Unlike metals, glass does not release trapped gases from its bulk at the temperatures of interest, so it does not need a deep bakeout; rather it requires only a thorough surface-layer scrub [3]. Consequently, we have bypassed this technological limitation through the highly non-trivial avoidance of all materials among the components entering the uninterrupted vacuum production line, other than glass (or a similar dielectric material) [1]. The simple ABALONE Photosensor design comprises only three industrially pre-fabricated (e.g. molded, pressed and/or cut) glass components (Fig.2 (c,d), Fig.3, Fig.4). These three glass pieces enter a continuous production-line process, which starts with standard rapid plasma cleaning (i.e. a surface scrub without a long bakeout), followed by a standard rapid thin-film deposition process, and ends with hermetic bonding of the three components, when the two thin-film seals between them are activated. Note that the two sealing surfaces are made of one of the deposited thin-films—the 'multi-functional smart alloy' invented in the UC Davis Ferenc-Lab in 2004 [4].

The low production cost is accompanied by simple integration methods, as shown in [2]. For example, ABALONE Photosensors may be integrated within thin and lightweight panels, serving as modules for ultrasensitive whole-body-enclosing (~20m$^2$ of sensitive area) medical PET scanners that provide ultra-low-dose cancer screening for a wider, even symptom-free population. The same invention enables custom-shaped detectors of illicit radioactive materials, or detectors for cosmic rays, cosmic gamma rays, neutrinos and dark matter. In all cases, these self-supporting detector shells can be assembled in the field from the encapsulated, vibration and shock proof, cable-free, electrically shielded and waterproof panels that seamlessly lock to each other [2]. Yet another advantage of the ABALONE Technology that stems directly from its simplicity is the low

---

[1] The signal-amplification process in these sensors goes via electron-hole pair multiplication, rather than photon multiplication. The effect of photo-multiplication nevertheless takes place as a side effect, and causes one of the main drawbacks of G-APDs – the photon crosstalk. The term 'photo multiplier tube' was an incorrect choice for the PMTs in first place.

level of radioactivity [1]. This is due to the optional ultrapure fused-silica composition of the components, as shown in the case discussed in this article and presented in Fig.4

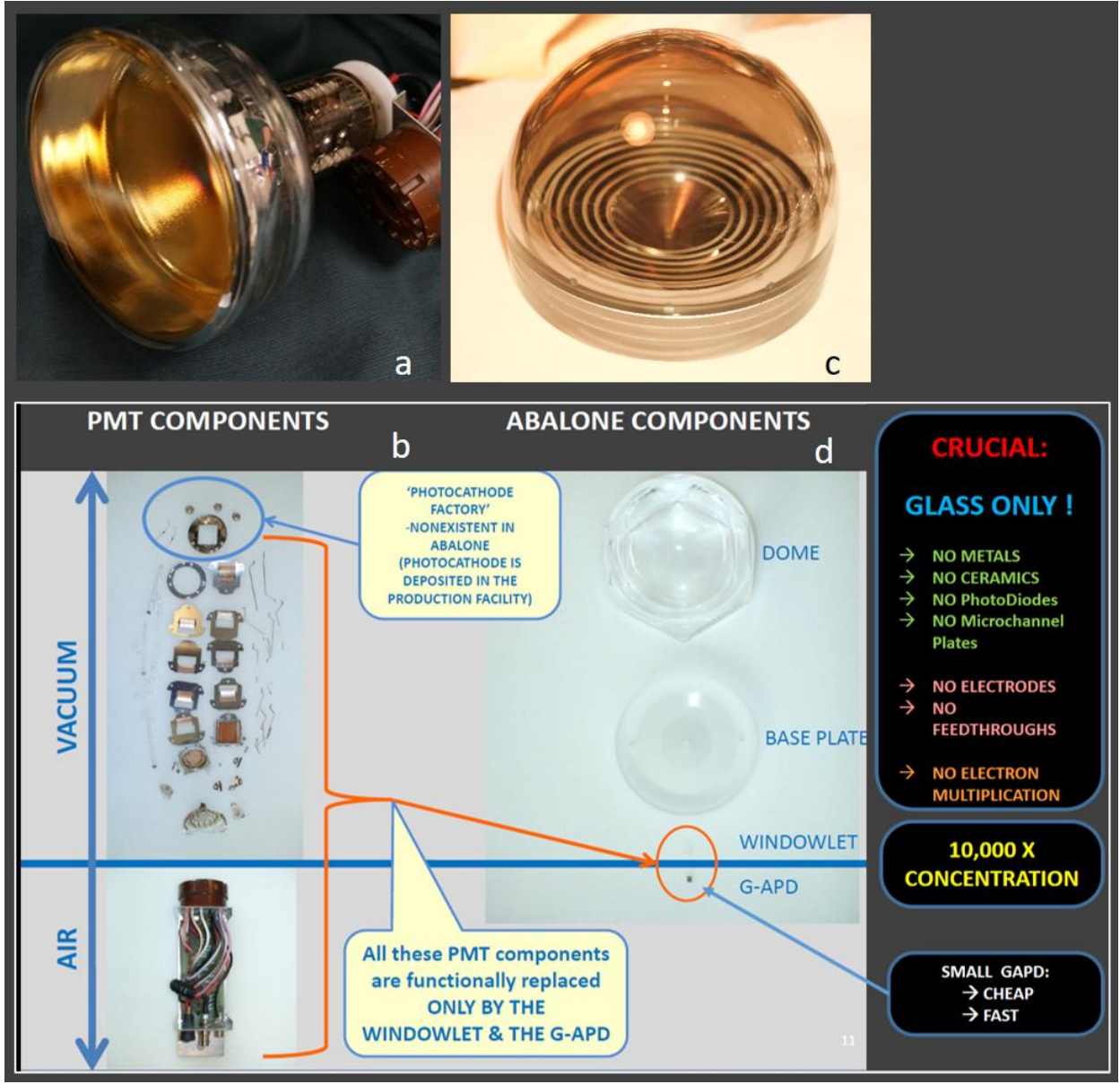

Fig.2 A PMT (a), and its components (b): more than 180 hand-made components, approximately 10 materials, more than 400 spot-welds (Fig.1); evaporators of photocathode-building materials remain in the PMT. The ABALONE Photosensor [1] (c) and its components (d): only 3 glass vacuum-processed components treated in a way equivalent to the production of blank CDs; ABALONE assembly process—patented mass-production technology. The ultimate simplicity leads to approximately two orders of magnitude lower production cost.

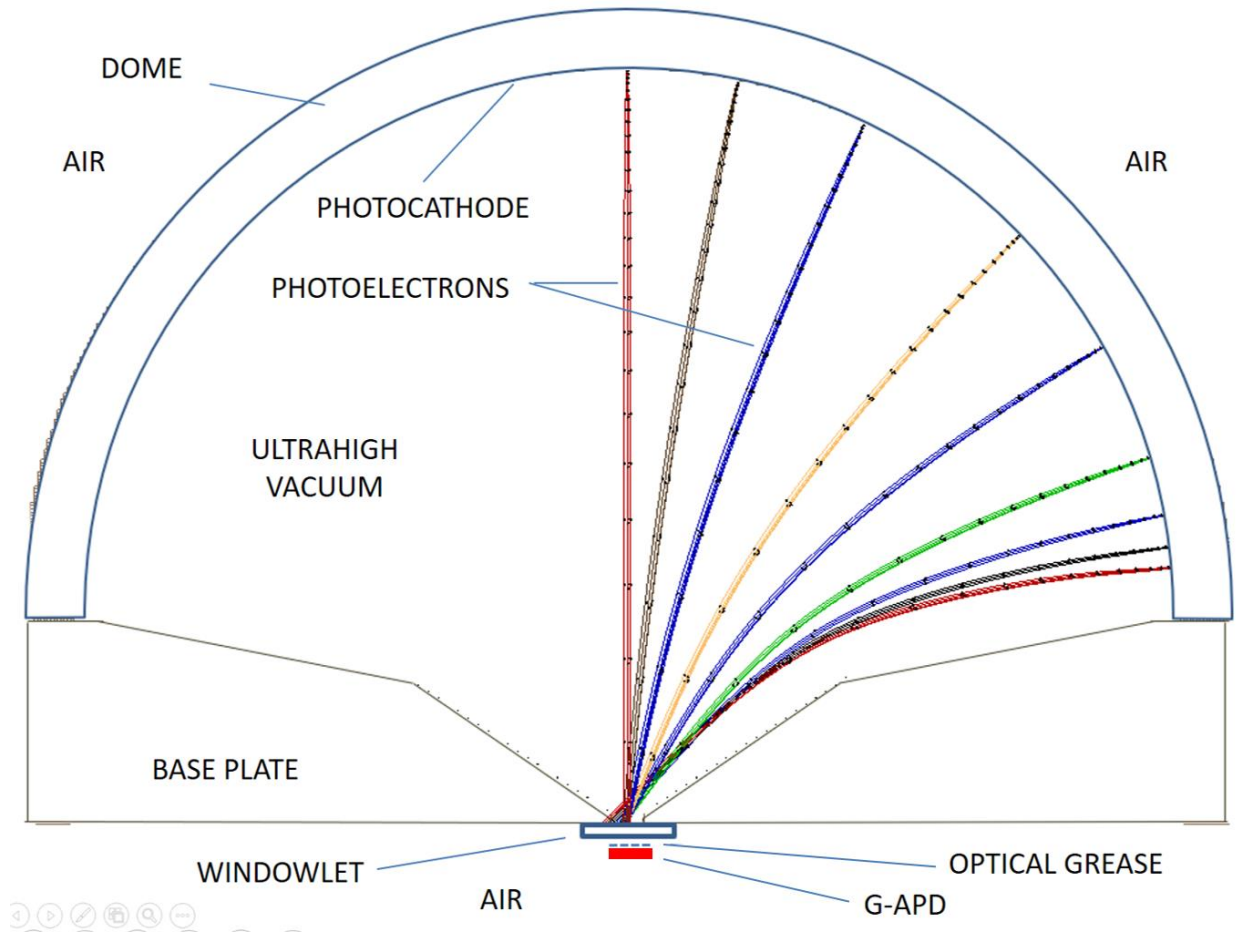

Fig. 3 Schematic view of the ABALONE Photosensor [1]; The electron trajectories result from the simulations of 20 keV electrons with a 0.25 eV transverse energy spread, for a Dome of a 10 cm inner diameter. The focusing mechanisms are explained in [1] and [2]. The concentration factor from the photocathode area to the readout area is ~10,000. A Geiger-Mode Avalanche PhotoDiode (G-APD) is optically coupled to the airside of the Windowlet to detect secondary photons created in impacts of photoelectrons in the Windowlet-scintillator.

ABALONE Photosensor prototypes have been produced in the specially constructed ABALONE Prototype Pilot Production Plant (A4P) (part of the Ferenc-Lab at UC Davis), which is in essence a downscaled rudimentary version of a real production line that nevertheless performs all of the processes accurately [1]. This facility was designed based on the R&D laboratory and the manufacturing equipment that was acquired from the Candescent company—the pioneer of flat-panel field emission TV technology [4], after its closure in 2001. That diverse and fully functional laboratory, along with the inherited know-how, has enabled the execution of our project. Just like Candescent, we were able to start with small-scale, fast-turnaround prototyping in the material research domain (2001-2004), followed by full-scale prototyping of isolated components and/or processes (2005-2013), and concluding with the assembly and stress-testing of functional prototypes (2011-2017). Some valuable manufacturing elements have also been adopted from a separately acquired Varo night-vision image intensifier production unit. Based on the results of the continuous R&D effort, PhotonLab, Inc. has recently designed and acquired a production-oriented facility, as well as a new testing and evaluation laboratory that has already been used in the hereby-presented analysis.

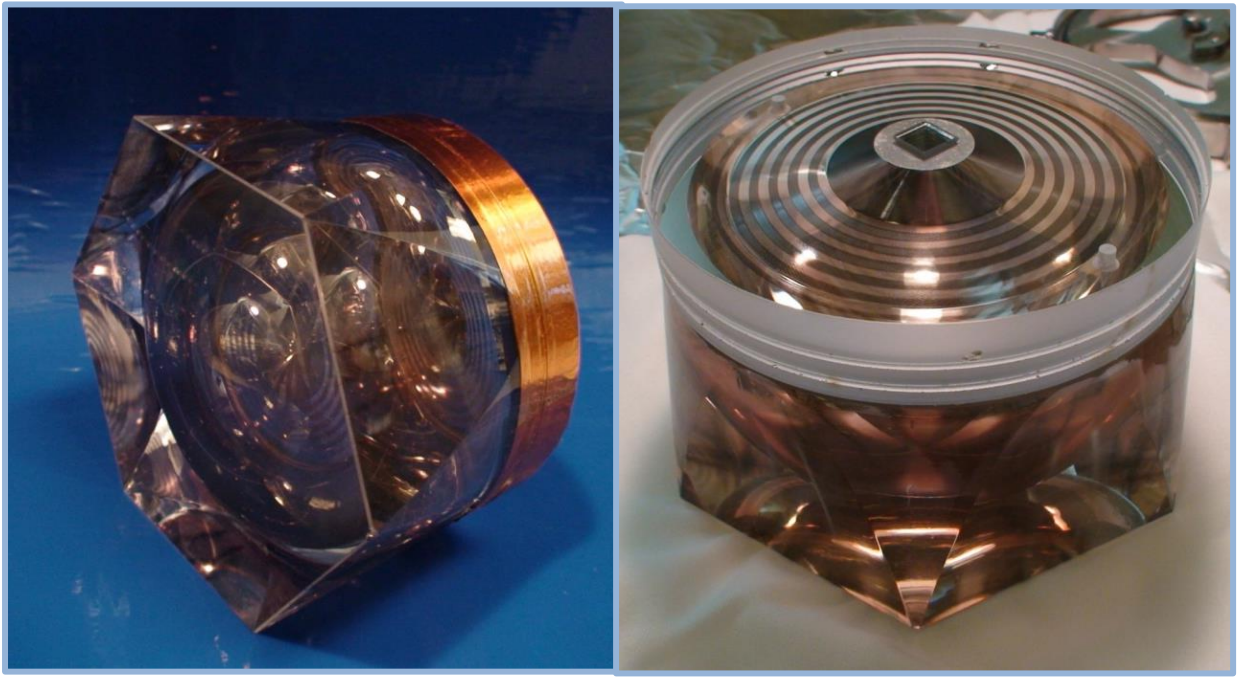

Fig. 4 The results presented in this publication are based on the tests of the shown ABALONE photosensor, continuously tested ever since its assembly in May, 2013.

One of the keys to this technology is our thin-film vacuum-sealing technique, optimized through several generations of full-scale prototypes whose components were recycled for repeated use dozens of times. Verifying the lasting vacuum integrity of a sealed vacuum enclosure is a complex task that can be properly carried out only over a long period of time. This article reports on the continuous, 4-year long test of one of the ABALONE Photosensor prototypes that was left working on the test bench ever since it was assembled in May of 2013. We report on the exceptionally low and constantly improving afterpulsing rate of positive ions, and explain the physics behind this result.

Further publications will report on the studies of other prototypes, and subjects including thermionic noise and performance at low temperatures, performance with a new generation of G-APDs and scintillators, the radioactive cleanliness for applications in dark matter search, modular panels, and the specially designed ABALONE Photosensors for large neutrino physics experiments.

1. THE *ABALONE* PHOTOSENSOR

A schematic view of the ABALONE Photosensor is shown in Fig.3. An ABALONE Photosensor comprises only three glass components that enter the assembly process. Two glass-to-glass bonds are established using our 'smart multi-functional thin-film,' one between the Dome and the Base Plate, and the other between the Base Plate and the Windowlet. Our novel bonding and vacuum encapsulation method [4] has been thoroughly tested and optimized ever since its conception in 2004. We have fine-tuned the latest generation of seals, focusing on the very small sealing surfaces between the Windowlet and the Base Plate as well as the mechanical strength of the seal. Destructive pulling tests have demonstrated that the seals never decouple; rather, the bulk glass material breaks.

These two conductive bonds play another key role: they pass the high voltage (between the Dome and the Base Plate) and the ground potential (between the Base Plate and the Windowlet) into the evacuated volume. This allows the ABALONE Photosensor to function without any through-the-glass feedthroughs, i.e. without any metals among the vacuum-processed components. As discussed, the absence of metals among the components is key to the ABALONE Technology.

Each photoelectron emerging from the photocathode on the inside surface of the Dome is focused and accelerated into the small hole in the center of the Base Plate. Practically all photoelectrons enter that hole, independently of the origination point, except for a small fraction ($\sim 10^{-4}$) of those that are significantly deflected in rare collisions with residual gas constituents. The photoelectron collection efficiency of ABALONE Photosensors is thus practically 100%. The hole in the Base Plate is covered and vacuum-sealed by the Windowlet (Fig.2 (d), Fig.3, Fig.4) from the outside. The concentration factor from the photocathode area to the bombarded Windowlet area is 10,000, which minimizes the G-APD area, and thus the overall cost. For instance, a detector of $\sim 1m^2$ area, composed of closely packed ABALONE Photosensors [2], would require only $\sim 1cm^2$ of total G-APD area. The small area of each individual G-APD directly translates into a small capacitance and resistance, i.e. it assures fast timing resolution.

On its vacuum-side, the Windowlet is coated with a multifunctional thin-film alloy, which: (i) provides a conductive coating over the photoelectron-receiving surface, (ii) connects that surface to the ground potential on the air-side of the photosensor, (iii) participates in the vacuum encapsulation process, in conjunction with the smart multifunctional thin-film coating, (iv) acts as a reflector, directing photons created in the scintillator towards the G-APD, and thus also (v) prevents secondary photons from reaching the photocathode. The Windowlet can either be entirely made of a suitable scintillator material, or as a thin glass plate coated on its vacuum-facing surface with a thin scintillator film (<2μm). The Windowlet-scintillator converts the kinetic energy of the accelerated photoelectrons into secondary photons, which are then detected by the G-APD.

When a photoelectron hits the surface of the Windowlet exposed to vacuum, it first penetrates the thin-film of alloy in which it loses a certain amount of energy. While electrons of 2.5 keV lose virtually all of their energy, electrons of 20 keV lose only about 0.3 keV, because the rate of energy loss in matter for electrons of these energies is inversely proportional to the kinetic energy [14]. The remaining energy allows each photoelectron to penetrate deep into the scintillator (<2μm), where it generates a large number of secondary photons, proportional to its kinetic energy and the scintillator yield.

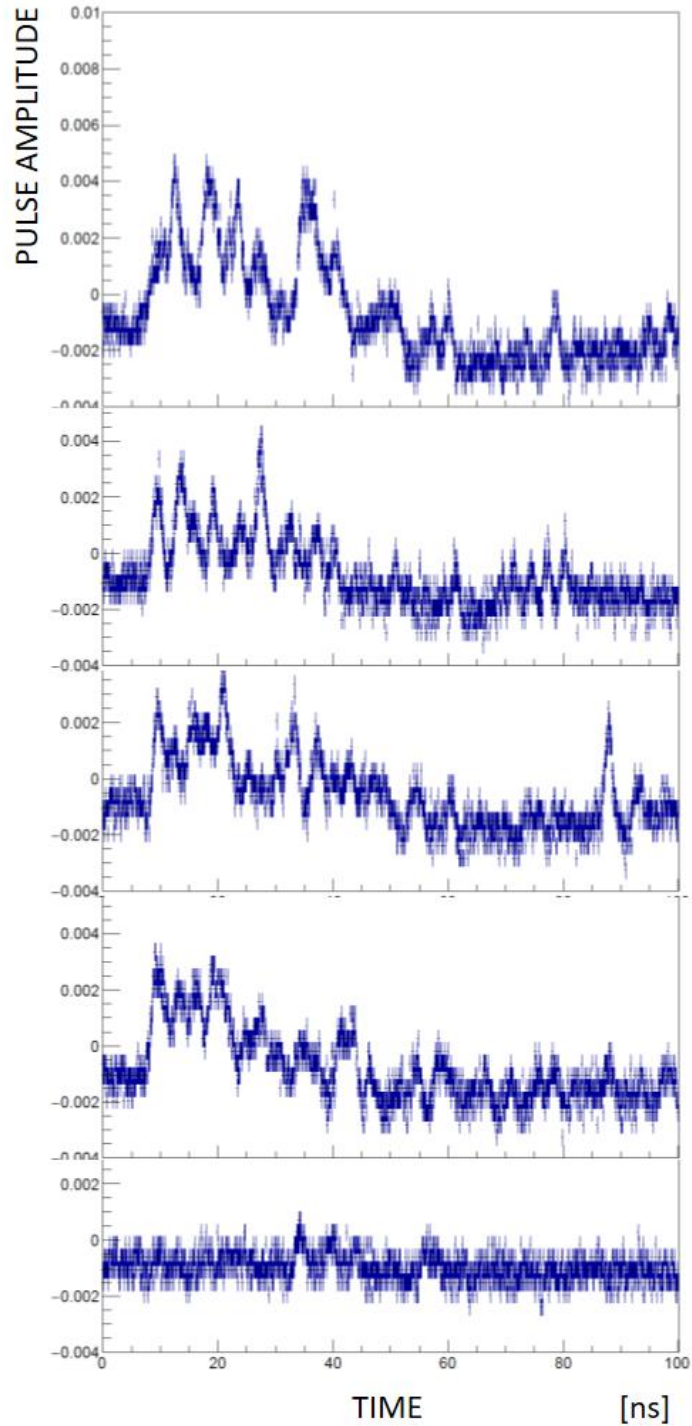

Fig.5 Examples of waveforms; fast signal output from the 3x3mm$^2$ SensL-J G-APD, directly recorded with the 5GHz oscilloscope at full bandwidth. These events correspond to 0, 1, 2, perhaps also 3 photoelectrons. Peaks due to the individual secondary photons from the Windowlet-scintillator are visible on the right sides. The decay time of the LYSO scintillator is 40-45ns, but the sharp leading edge of each pulse allows for sub-ns time resolution.

The ABALONE Technology creates a class of its own. Regarding the light detection principle, ABALONE Photosensors just vaguely belong to the class of Image Intensifiers. More accurately, they belong to a separate class, which we have previously defined as Light Amplifiers [15,16] in order to underline the key distinctive feature: light amplification, but without imaging. The other two distinguishing features are the response time (scintillator vs. phosphor) and the specific method of detection of the secondary light (G-APD vs. CCD, or the naked eye). However, a G-APD in an ABALONE Photosensor can be replaced with other sensors, such as an analog APD, a PiN diode, a miniature CCD, or even the naked eye. We have used the last two methods in stress tests (2012) of yet another ABALONE Photosensor prototype, by exposing it to strong sources of light, including even daylight. Indeed, a bright spot on the air-side of the Windowlet was clearly visible. Another fused-silica prototype was immersed in liquid nitrogen, without damage. It is worth stressing that ABALONE Photosensors are virtually resistant to shock and vibration, since the three components form a strongly bonded unit. The absence of light-induced damage, along with manifest mechanical robustness and overall simplicity, makes them sturdy and safe.

## 2. EXPERIMENTAL

The Dome of the ABALONE Photosensor used in the hereby presented tests (see Fig.4) has a hexagonal shape on the outside (an approximation to the Winston cone for close packaging into panels [2]), and is made of fused silica. A more recent hemispherical prototype, made of an optical glass, is shown in Fig.2(c). Any bialkali or multialkali photocathode can be grown in the Dome under controlled conditions within the production system, controlled by a quartz crystal thickness monitor, optical transparency, and photocurrent response.

For this particular ABALONE prototype, we chose the $Cs_3Sb$ photocathode, because it spontaneously converges to the optimal stoichiometric configuration and standard quantum efficiency (for a given thickness and substrate) [9][10][11]. This makes it ideal for R&D projects like ours that focus on topics other than the photocathode. However, this photocathode material has been of a particular interest also because of its demonstrated potential for fast industrial production [12]. Our photocathode has maintained an expected efficiency of approximately 8-13% at 405nm [11][10], as estimated through an analogy to the response of a G-APD illuminated under the same conditions. Note that some other photocathode materials may also feature a high level of manufacturability and robustness [13].

The Windowlet is a 6x6x1.5mm$^3$ plate made of the LYSO scintillator material (Cerium-doped Lutetium Yttrium Orthosilicate). We chose LYSO for this study because of its mechanical and chemical robustness, rather than for its yield or speed. A family of other scintillators that offer higher yields and faster response than LYSO have been studied recently, but an ABALONE Photosensor with a LYSO scintillator already provides an acceptable performance for most applications. The optical coupling of the G-APD to the Windowlet was done with standard optical grease rather than a permanent optical cement, at the expense of light-transmission efficiency, because that allowed us to test so far seven different G-APDs from different manufacturers. The possibility of replacing G-APDs at any time is yet another important intrinsic advantage of the

ABALONE concept. The most recent data presented in this article (2015-2016) were collected with a 3x3mm$^2$ SensL J-type MicroFJ G-APD that comprises 35μm cells, and the older presented data were taken with several Hamamatsu G-APD models from 2012-2014.

A 23keV photoelectron generates an estimated 600 photons in the LYSO scintillator. Approximately 40-100 of them get recorded, depending on the G-APD model and bias voltage. The latter estimate came from precise counting of the individual photons, resolvable in the fast signals (see Fig.5) of the SensL J-type G-APD (to be published separately). Our 1-photoelectron signal therefore dwarfs the G-APD noise, and guarantees excellent single-photon sensitivity and resolution.

We applied a negative, ultra-low current high-voltage to the photocathode, while the Windowlet and the attached G-APD were shielded within a miniature grounded conductive cylinder that forms a Faraday cage together with the Windowlet and the sealing surface around it. The ABALONE Photosensor consumes practically no power. A high-voltage power supply similar to the miniature supplies in night-vision goggles can be used, but still of a significantly lower power. The acceleration potential can range between approximately 4kV and 25kV (successful tests went up to 30kV), depending on the application. A 405 nm LED light source, centered axially 42cm above the prototype, was pulsed at 100 kHz by an SRS DG535 Pulse Generator. We have used two separate systems for data acquisition. A standard NIM setup (based on Ortec, Canberra, Tennelec and LeCroy components, including fast amplifiers, pulse shapers, coincidence units, discriminators, time-to-digital converters, and an MCA) has been used before 2015, and more recently a pair of LeCroy digital oscilloscopes (1GHz and 5GHz) has been used almost exclusively. The data were analyzed using the CERN-Root software package.

3. RESULTS

The presence of residual gas in any photocathode-based vacuum photosensor is critical, because it both leads to the ion-feedback afterpulsing noise and harms the photocathode and other surfaces due to ion impacts and chemical contamination. This article reports on the exceptionally low and constantly improving afterpulsing rate of positive ions in the tested ABALONE Photosensor, and explains the physics and technology behind this result.

We first discuss the diagnostic tool that we have used to constantly monitor the vacuum quality and the chemical composition of residual gas. An ABALONE photosensor functions as a time-of-flight (TOF) mass spectrometer for ions generated within its own vacuum. A fraction of approximately 10$^{-4}$ of all photoelectrons in the ABALONE Photosensor ionize a neutral atom or a molecule on their way towards the Windowlet, or on the Windowlet surface itself. Each of those ions is then accelerated towards the photocathode, where secondary electrons can emerge upon its impact. Those electrons are detected the same way as photoelectrons, unless they miss the Windowlet due to an excessive initial transverse momentum acquired during the ion impact. Their detection corresponds to the after-pulse, i.e. the "stop" signal in the ion TOF measurement, while the "start" signal comes from the detection of one or more of the initial photoelectrons (Fig.6). The time of flight of an ion is the time difference between the "stop" and the "start" signals, corrected for the electron's own flight time.

The TOF spectra shown in Fig.7 (taken in November of 2016), and Fig.8 (taken in January of 2014, i.e. 7 months after the production of the prototype) feature peaks corresponding to ions of different masses and charge states. The spectrum in Fig.8 is shown in logarithmic scale to emphasize rare components. We have fitted a sum of Gaussian functions to TOF spectra, where each Gaussian function corresponds to a light ion typically present in vacuum. We set the initial TOF values to the results of numerical simulations, assuming for simplicity that all ions originate from the Windowlet surface. For most atoms and molecules other than noble gases, that assumption is partly true, because they are in thermal equilibrium between their free and adsorbed states.

The data shown in Fig.7 were taken directly with an oscilloscope, without amplification, which has allowed precise pulse separation at short time intervals. We have used the fast output from the SensL G-APD in this analysis, bandwidth-limited to 20MHz, filtered with $\sin(x)/x$ and LeCroy's ERES3 interpolation that provides an upscale from 8 to 11 bit resolution. The downside of this method is a poor single-photon resolution (Fig.9) due to a simple pulse height readout that takes only the peak value, neglecting the baseline and the pulse shape (Fig.6). For completeness, in Fig.10 we show a properly extracted single-photon resolution spectrum (taken at a slightly lower light intensity setting). The data in Fig.8 were taken with NIM electronics that has shaped and stretched the pulses, and brought the 'start' and the 'stop' peaks into overlap, causing a cutoff in the TOF spectrum below 150ps.

The initially observed traces of C, N, O, CO/$N_2$, $CO_2$, $H_2O$ and Ar, still present in Fig.8, have been falling since the beginning of our continuous test in May of 2013. However, the presence of He has remained constant, consistent with the fact that for any vacuum enclosure exposed to air, helium penetrates glass until its pressure reaches its partial atmospheric pressure. The rate of molecular ions has dropped most rapidly due to molecular dissociation.

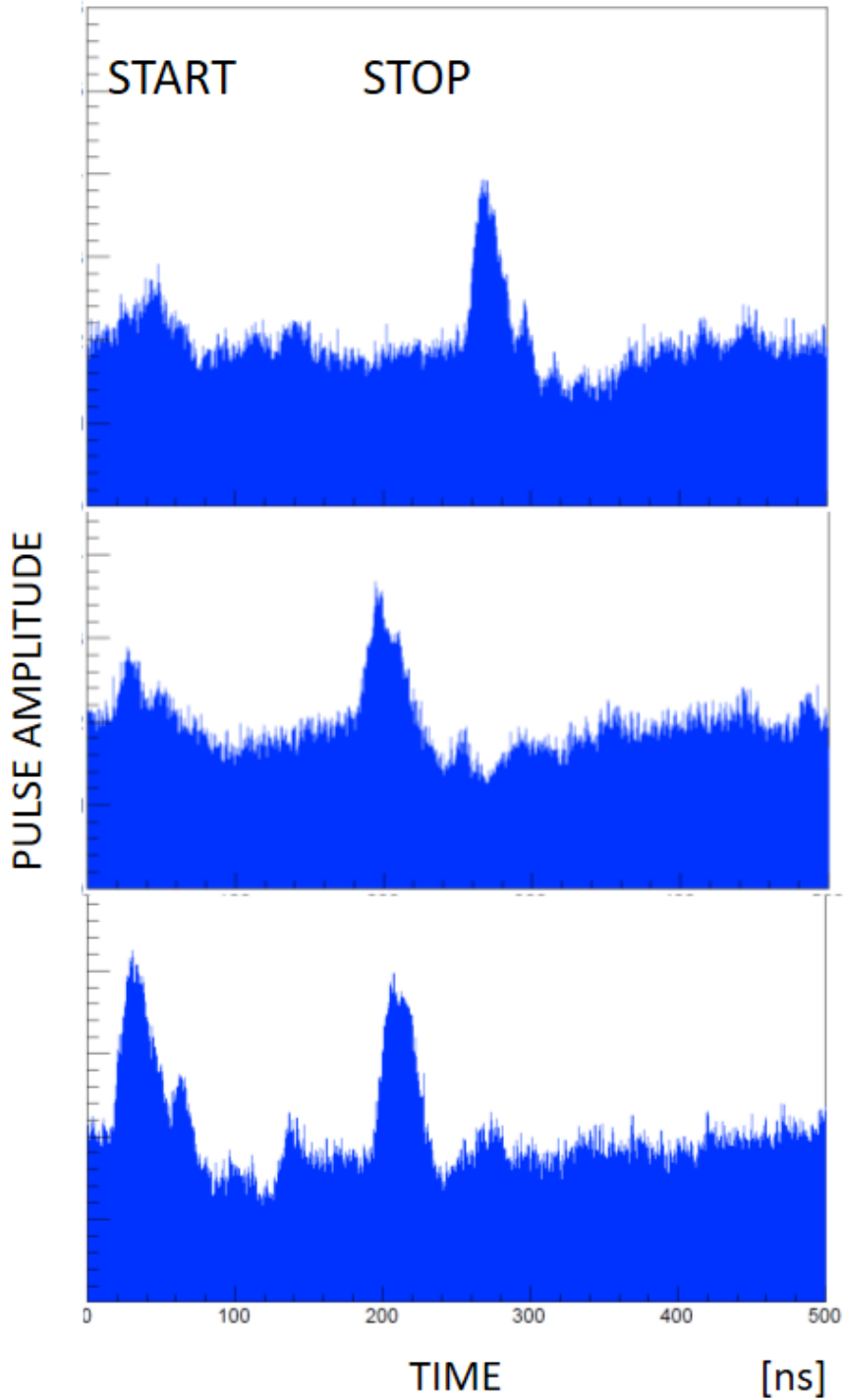

Fig.6 Examples of events triggered by the existence of a 'stop' after-pulse following a 'start' pulse caused by the detection of photons from the LED pulser. The presented waveforms are the fast signals from the 3x3mm² SensL-J G-APD, bandwidth-limited to 20MHz and filtered by the oscilloscope. The period between the 'stop' and the 'start' peaks is the time of flight of an ion (extended by electrons' flight time).

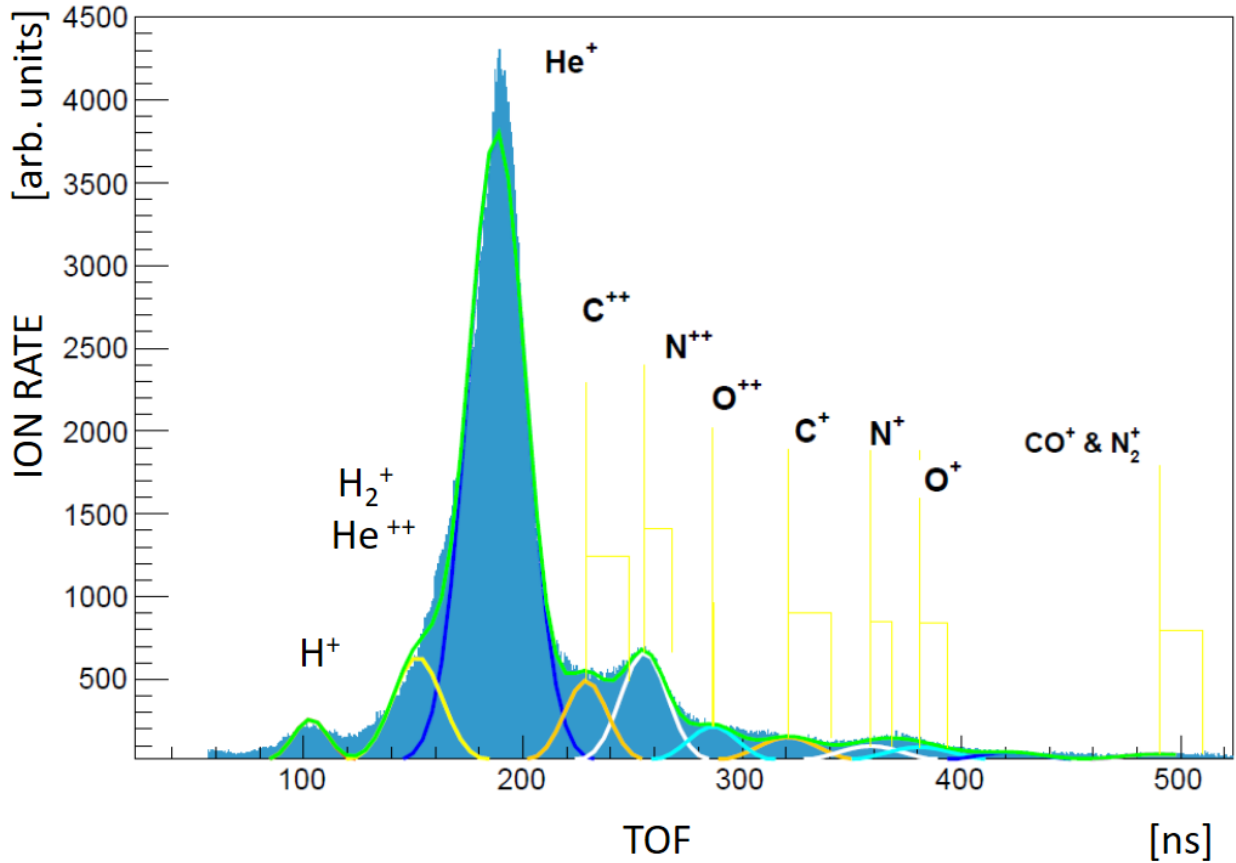

Fig. 7 Time-of-flight mass spectrum of residual gas within the ABALONE Photosensor, taken in November of 2016. Acceleration potential of 20kV, the threshold on after-pulse amplitude was at approximately 1.5-2 photoelectrons. Short vertical lines indicate the numerically predicted TOF values assuming that the ionization is taking place on the surface of the Windowlet, which served as the initial values in the fitting. Long vertical lines correspond to the best overall fit.

The total afterpulsing yield is the integral of the mass spectrum. Since the afterpulsing rate drops rapidly with the after-pulse amplitude (see Fig.9), the precision of the 'stop' signal's amplitude measurement, as well as of the threshold setting strongly influences the afterpulsing rate. That makes the comparison of the afterpulsing rates of different sensors rather difficult, particularly when the threshold is set to a high after-pulse amplitude.

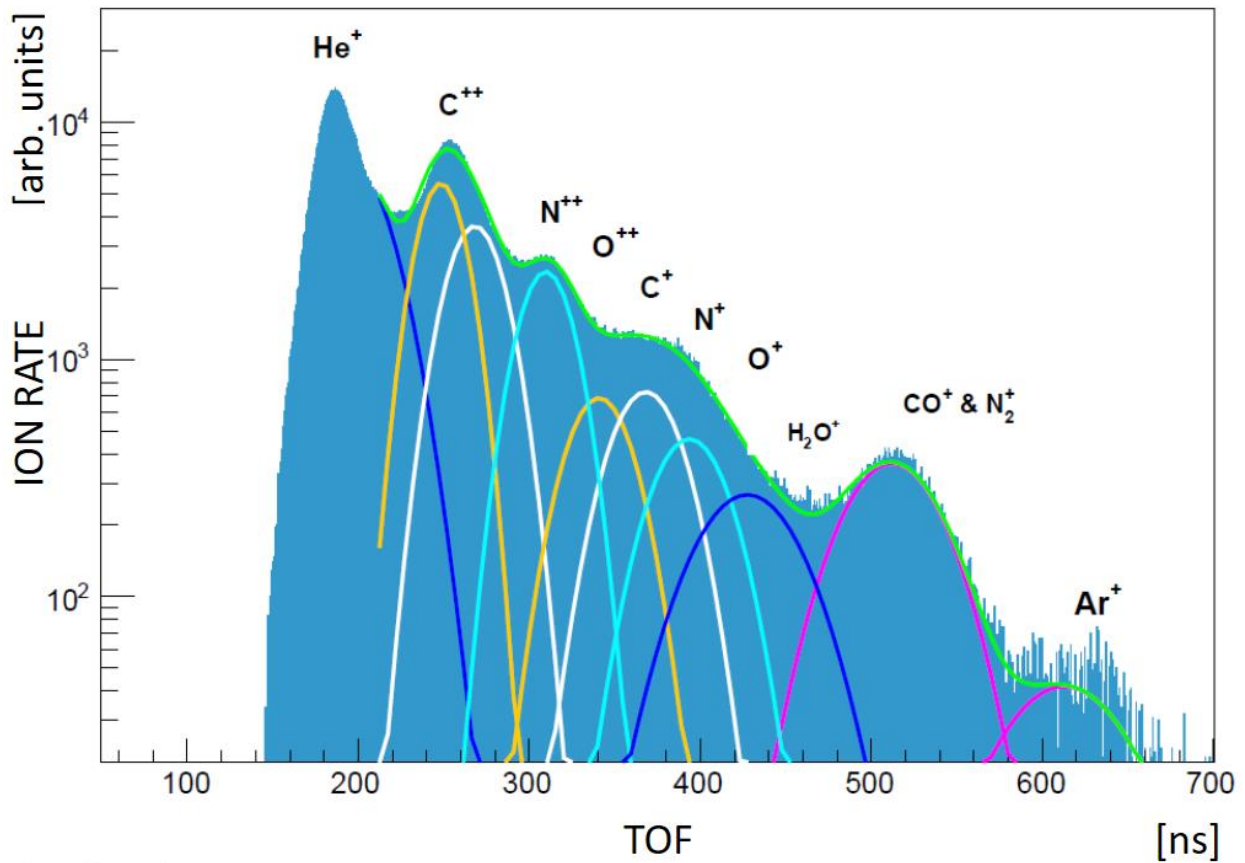

Fig.8 TOF mass spectrum taken in January of 2014, when molecules and atoms other than He were still relatively abundant. Acceleration potential of 20kV.

Another uncertainty lies in the normalization of the afterpulsing yield to the average number of photoelectrons. The average number of photoelectrons that contribute to an observable afterpulsing event is higher than the average number of photoelectrons, because a single photoelectron is likely to be lost in the ionization process and may not give a 'start' pulse. That shifts the average number of electrons higher, i.e. the normalized afterpulsing rate lower. For all these reasons, we have repeated the same measurement three times with slightly different settings, and a systematic error of 50% has emerged. Furthermore, it may be misleading to compare new sensors with long-term stress-tested ones, like ours. Contrary to ABALONE, the vacuum within a typical PMT usually deteriorates with time and usage. In any comparison of this kind, one should also take into account the path traveled by photoelectrons within the sensor, i.e. one should compare devices of similar dimensions. However, some ions originate from the impacts of photoelectrons on the surfaces. In spite of all of these difficulties, we have estimated the afterpulsing rates in conditions similar to the recently tested, new, and approximately 3 times smaller PMTs [8].

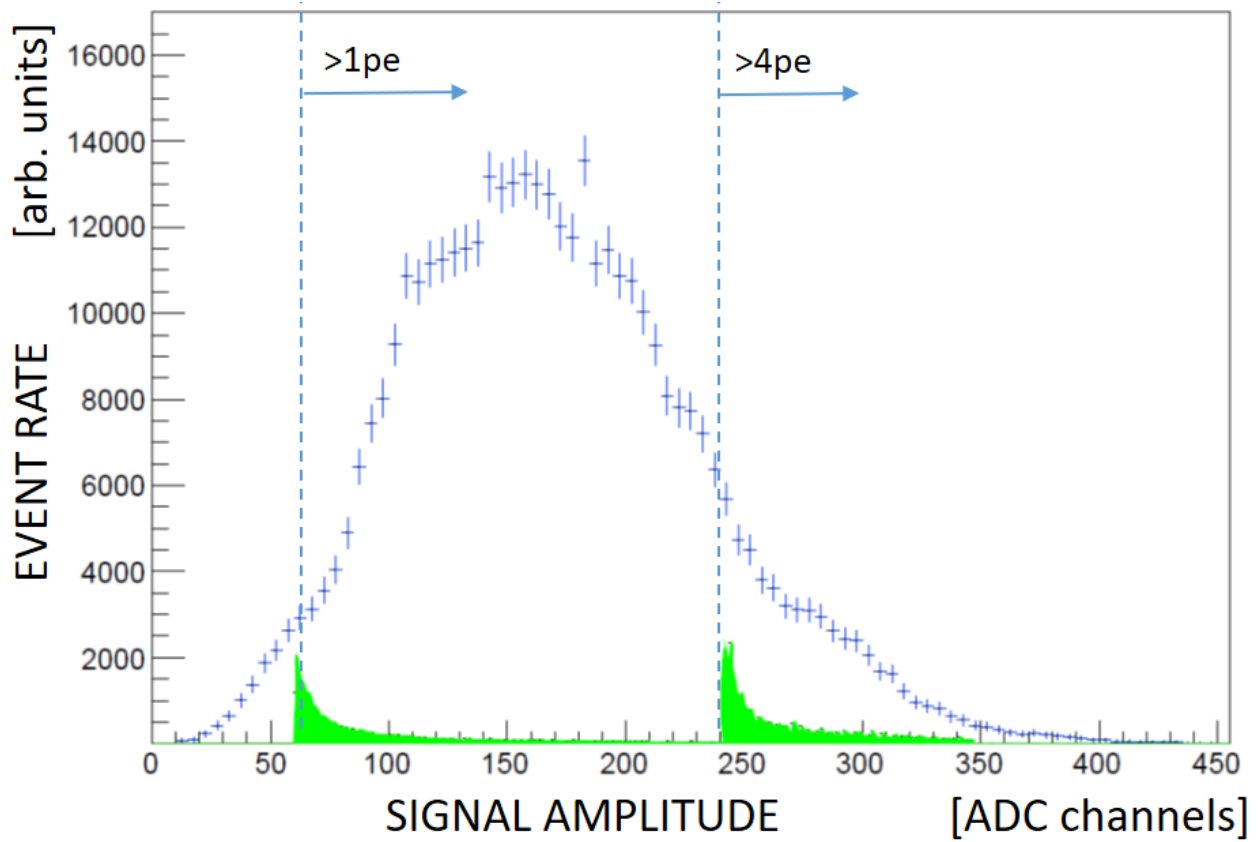

Fig.9 The pulseheight spectrum of 'start' signals. The TOF mass spectra in Fig.11 were taken with threshold levels indicated in this figure. The two corresponding 'stop' or after-pulse electron spectra (not to scale) are shown in green at the bottom of the figure. Acceleration potential 22kV.

The amplitude spectrum of "start" pules is shown in Fig.9, together with two 'stop' after-pulse amplitude spectra (not to scale), corresponding to the two thresholds. The 4-electron threshold was set at the fourfold single-photon peak amplitude. Fig.11 shows the two TOF mass spectra, obtained with the corresponding two amplitude thresholds. As expected [18][19], heavier and/or double-charged ions contribute more to afterpulsing events of higher amplitudes, while protons and alpha particles penetrate deeply, with a very low loss of energy at the surface.

We estimated the afterpulsing rates as $(8+/-4) \times 10^{-4}$, and $(7+/-5) \times 10^{-5}$ per photoelectron, for after-pulses of amplitudes higher than 1 photon, and 4 photons, respectively. That is approximately two orders of magnitude lower than the rates reported in [8] (taking into account the factor of 3 difference in the sizes). As discussed below, this result should not be surprising, given the fundamental differences in the operation principles and production technologies.

It may seem puzzling that an ABALONE Photosensor features an afterpulsing rate that is approximately two orders of magnitude lower than that of PMTs of the same size, while the helium pressure should arguably be the same in both vacua. However, one should recall that ABALONE uses two orders of magnitude higher potential to accelerate photoelectrons, and that the ionization stopping power is inversely proportional to electron's kinetic energy [14]. Thus, due to their two orders of magnitude higher energies, photoelectrons in ABALONE simply ionize two orders of magnitude less helium than the photoelectrons in PMTs[2]. This fundamental advantage also holds for the other constituents of residual gas.

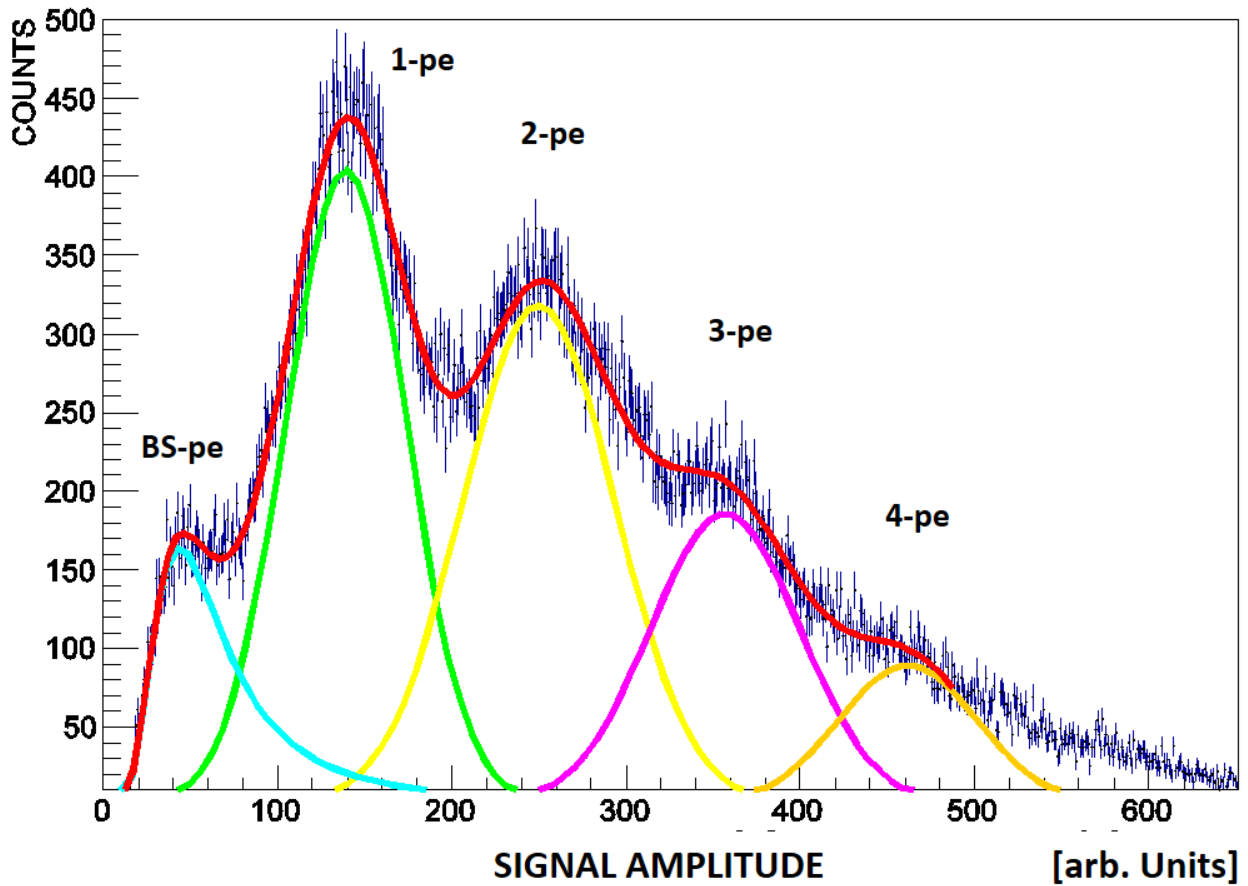

Fig.10 A typical amplitude spectrum taken at an acceleration potential of 23kV, with a slow signal of the 3x3mm$^2$ SensL-J G-APD, at a 28.5V bias voltage. Gaussian function components in the common fit represent photoelectron counts from 1 to 4, while the Landau function (BS-pe) approximates the contribution of non-returning back-scattered single photoelectrons that leave a measurable fraction of their energies in the Windowlet, before backscattering into the vacuum ($\chi^2$/ndf=532/464). Backscattered electrons within multi-photoelectron events fill the gaps between the individual photoelectron peaks. Some backscattered photoelectrons return into the Windowlet after looping through vacuum, so they are detected with their full amplitudes.

---

[2] Note that ion gages and mass spectrometers, which benefit from maximizing the ionization probability, use electrons of practically the same energy as PMTs.

The exceptional level of vacuum cleanliness comes because of the following key factors:
- The intrinsic cleanliness of the ABALONE Photosensor Technology (e.g. base pressure of $10^{-11}$ Torr in the sealing section of our assembly line)
- The absence of any outgassing sources, such as metals, ceramics, feedthroughs, microchannel plates, silicon diodes
- Complete absence of electron multiplication, and thus electron-induced desorption within the vacuum
- The integrity of the two thin-film seals
- Minimum area, open geometry and high molecular conductivity
- The three internal vacuum pumping mechanisms that constantly improve the vacuum level in ABALONE:
    - Ion implantation (explained below),
    - Physisorption on passive surfaces (none of the surfaces within ABALONE, except of the small electron-receiving Windowlet area, suffer from any electron bombardment),
    - Chemisorption on the multi-functional, chemically reactive thin getter film that covers 90% of the projected Base Plate area.

In general, the interaction of an energized ion with a solid depends on the properties of the solid, ion's energy, its atomic number, charge, and the angle of incidence. For light elements (from H to O), the stopping power comes as a superposition of the so-called 'nuclear' interaction (nucleus-nucleus Coulomb scattering), and the low-energy tail ($dE/dx \sim E^{1/2}$) of the 'electronic' stopping power peak (excitation and ionization of atoms) [18, 19]. At 20keV, a light ion experiences a rather low stopping power (~3-20eV/atomic layer, from H to Ar), so it easily passes through the thin photocathode film (~15nm) with enough energy left to penetrate deep into glass (~50-200nm), where it stops[3]. Once implanted in glass, practically all atoms except helium remain trapped, never returning into vacuum. Ion implantation thus acts as a very efficient vacuum pump. For instance, we have observed traces of argon disappearing soon after we started with this test, see Fig.8. As a noble gas, argon may not be chemisorbed in a getter, and therefore ion implantation remains the only possible explanation for its removal.

In contrast to ABALONE, ions generated in PMTs gain energies of only up to about 200eV. The stopping power for ions below 500eV in any solid is so high that they come to an abrupt stop right on the surface. Hard 'nuclear' collisions dominate and the displacement of target atoms is common[4]. The entire ion's energy is thus transmitted in one or just a few collisions with atoms in the photocathode surface. Some ions may stay chemisorbed, contaminating the critical surface layer of the photocathode, while the others neutralize and return into vacuum. Vacuum pumping through ion-implantation in glass is impossible in PMTs and MCP-PMTs. Furthermore, in those sensors the electron-induced desorption may be by a factor of $10^5$-$10^6$ more frequent than in an

---

[3] Ion implantation is a widespread industrial technique used for doping of solids deeply under the surface, while making minimal damage to the surface and a number of sub-surface layers. The optimal ion energies for that process include the range of ion energies used with ABALONE.

[4] To maximize the sputtering efficiency of argon ions in a plasma, thin-film deposition systems typically bias sputtering sources to a potential of -500V.

ABALONE photosenor, since, by design, each electron participating in the multiplication process bombards a solid surface with energies of high ionization probability.

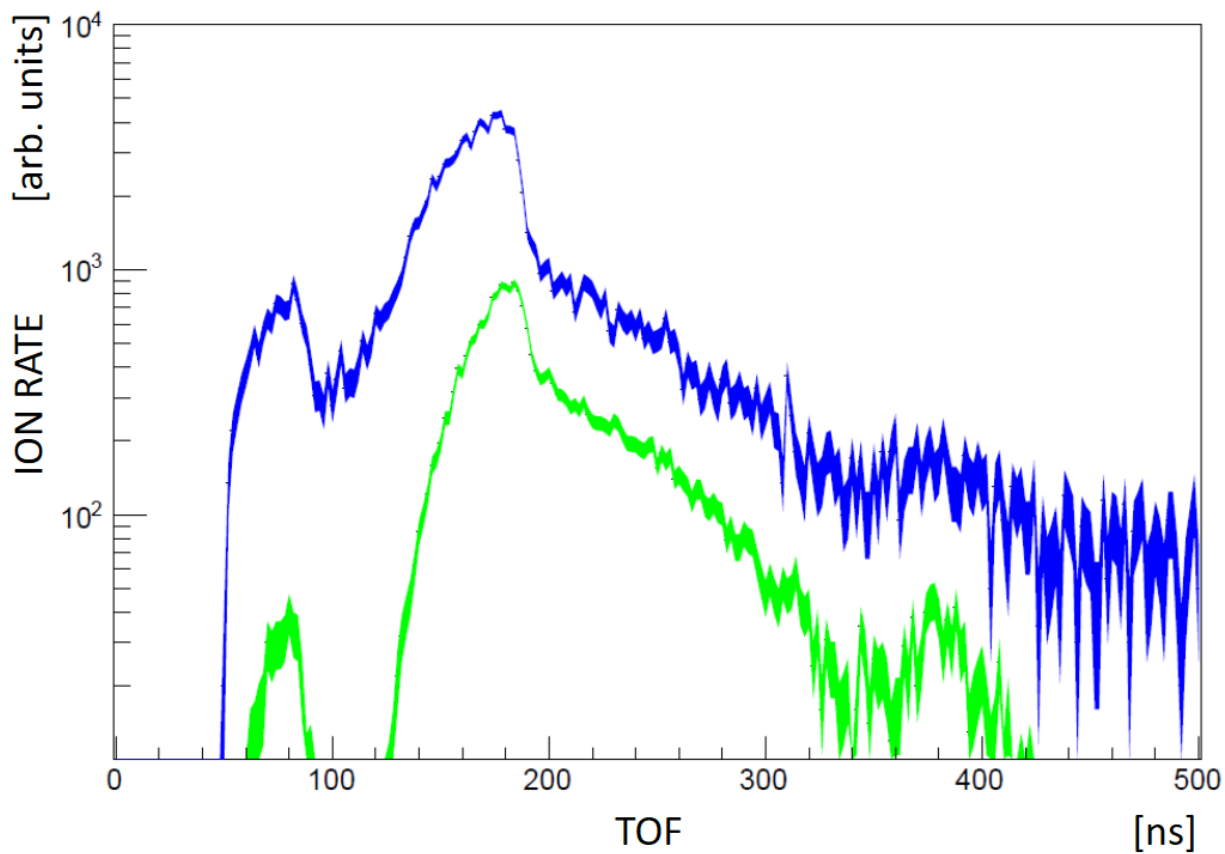

Fig.11 TOF mass spectra taken with two thresholds on the after-pulse amplitude, as indicated in Fig.9. The lower curve corresponds to the threshold at 4 electrons. The total rate of after-pulses of amplitudes above 4 electrons is approximately $10^{-5}$ per photoelectron. Acceleration potential 22kV. The presented data were taken in 2016, when most of the molecules, still present in Fig.8, have already dissociated. After-pulses of higher amplitudes (lower spectrum) are more likely to originate from heavier and/or double-charged ions. Protons (the peaks at the left) and alpha particles (the left shoulder of the dominant peaks), see also Fig.7, feature a particularly low rate of ionization on the surface.

4. SUMMARY

The exceptionally simple ABALONE Photosensor comprises only three glass components entering the vacuum process. Through the absence of any component material other than glass (metals, wire feedthroughs, ceramics, brazes, photodiodes, micro-channel plates, spot-welds), this invention has bypassed the technological limitation that has restrained the 80-year old photomultiplier tube (PMT) manufacture, as well as other similar vacuum photosensor concepts.

The ABALONE Technology involves only standard mass-production techniques (rapid plasma cleaning, thin-film deposition, robotic transport and positioning), all performed in an uninterrupted vacuum production line. This technology is in essence equivalent to the production of blank CDs.

In order to verify our concept over a long operation period, in 2013 we spared one of the early ABALONE Photosensor prototypes from reuse and kept it on a test bench for continuous stress testing. We report on the exceptionally low and constantly improving ion afterpulsing rate (approximately two orders of magnitude lower than in PMTs), and we explain the physical and technological reasons for this achievement.

Thanks to the low cost (approximately two orders of magnitude lower than that of PMTs, including integration and electronics) and the specific combination of properties, including low level of radioactivity, integration into large-area panels, and robustness, this technology can open new horizons in the fields of fundamental physics, functional medical imaging, and nuclear security.

ACKNOWLEDGEMENTS

This project was partly supported by the University of California Office of the President, under the Program UC Proof of Concept Award, Grant ID No. 247028 (2013-2014).
The authors thank technicians David Hemer and Keith Delong for devoted and professional work on this project.
The authors also thank Bruno and Dan Ferenc Šegedin for writing assistance and language help.